\documentclass[10pt,a4paper]{article}

\usepackage{amscd} \usepackage{amsmath} \usepackage{amsfonts}
\usepackage{epic}
\usepackage{latexsym}

\def\be{\begin{equation}}
\def\ee{\end{equation}}
\def\bea{\begin{eqnarray}}
\def\eea{\end{eqnarray}}

\def\A {$AdS_3$}
\def\AA {$AdS_2$}
\def \SL {$SL_2(\mathbb{R})$}

\def \H {$H_3^+$}

\def \p {\partial}

\def \tq {\tilde{q}}
\def \Z {\mathbb{Z}}

\def \R {\mathbb{R}}

\def \half {\frac{1}{2}}

\def \nn {\nonumber}
\def \llangle {\left\langle}
\def \rrangle {\right\rangle}
\def \bz {\bar{z}}
\def \j {-\half +iP}

\def \rr {\frac{r}{\pi b^2}}
\def \halfi {\frac{i}{2}}
\def \ip {i\frac{\pi}{2}}
\def \halfib {\frac{i}{2b^2}}

\title{{\small \begin{flushright}
CPHT-RR 071.0902 \\ \vskip .3cm
\end{flushright}} Two \AA\ branes in the Euclidean \A}
\author{Sylvain Ribault\footnote{ribault@cpht.polytechnique.fr }\\
{\it\small  Centre de Physique Th{\'e}orique, Ecole Polytechnique,
91128 Palaiseau, FRANCE}}
\date{}

\begin{document}

\maketitle

\centerline{ \bf Abstract}

\vskip .1cm

We compute the density of open strings stretching between \AA\ branes
in the Euclidean \A. This is done by solving the 
factorization constraint of a degenerate boundary field, and the
result is checked by a Cardy-type computation. We mention applications
to branes in the Minkowskian \A\ and its cigar coset.

\vskip .5cm

The \AA\ D-branes in \A\ have received some attention after the work
of Bachas and Petropoulos \cite{bp}. After some semi-classical studies
\cite{pr,plot}, important progress was made with the determination of
exact boundary states \cite{pst,pol} and open-string spectra
\cite{pst} for the
\AA\ branes in the Euclidean \A. This progress relies on the
understanding of closed string theory in \A\ and its Euclidean version
\H\ \cite{moi,te}.

However, the determination of the boundary states was done by solving
the factorization constraint of only one degenerate bulk field. Their
consistency is thus far from being proven
\cite{pon}. Another problem appeared in \cite{pst}, where it was argued that a
consistent density of state for open strings stretched
between two different \AA\ branes could not be found by solving the
boundary factorization constraints. It would be very strange
that each individual \AA\ brane would be physical, but that one could
not consistently stretch open strings between them.

By examining carefully the analyticity properties of the density of state, I
will argue that it is in fact possible to find a solution to the
boundary factorization constraint for two different \AA\ branes. The
result can be found in eqs. (\ref{oppo}) and (\ref{any}) (written in
terms of reflection amplitudes; I will explain their relation with
open-string densities). 

This result is not only reassuring for the
consistency of the \AA\ branes in \H, but also has implications for
branes in \A\ and the cigar $SL(2,\R)/U(1)$. Indeed, strings
stretching between two opposite \AA\ branes are related by spectral
flow to strings with winding one-half. Thus, our result (\ref{oppo}) gives
the density of long strings with odd winding number living
on a single \AA\ brane in $AdS_3$. As predicted in \cite{plot}, this
differs from  the density of long strings with even winding
number. Moreover, the \AA\ branes descend to D1 branes in the cigar,
which have open string modes with half-integer winding. Their density
is also given by eq. (\ref{oppo}), see \cite{prep}. This was in fact the
original motivation for this work. 

\vskip .2cm

The plan is as follows. First the density of states $N(j|r_1,r_2)$ and
reflection amplitude $R(j|r_1,r_2)$ for open strings between two
$AdS_2$ branes are introduced. The constraint on $R(j|r_1,r_2)$ is
written (\ref{good}) and the solution given explicitly
(\ref{same})-(\ref{any}). These formulas, and the sense in which they
solve (\ref{good}), are then made precise by a discussion of the
choice of the branch cuts in the
function $S'_k(x)$ (\ref{spk}). These results on the open string
spectrum are checked by a Cardy-like computation using the
boundary state (\ref{etatbordphijnp}). The note ends with a brief
explanation of why the spectrum of open strings between two different
$AdS_2$ branes was not found in \cite{pst}.

\vskip .4cm

Now let me define precisely the quantities I want to compute and the
notations I will use.
The spectrum of strings in \H\ is made of continuous representations
of \SL\ with a spin $j=-\half+iP \in -\half +i\R$ and a Casimir
$-j(j+1)$. All these states behave asymptotically as plane waves. 
If we write $ds^2=d\psi^2+\cosh^2\psi(e
^{2\chi}d\nu^2+d\chi^2)$ the metric of \H, then we will be interested
in \AA\ branes of equations $\psi=r$ for some real constant $r$. 

The spectrum of strings stretching between two such branes $\psi=r_1$
and $\psi=r_2$ can
be described by a density of states
$N(j|r_1,r_2)=N(P|r_1,r_2)$. This is linked to another physical
quantity, the  
reflection amplitude $R(j|r_1,r_2)$ such that
$N(j|r_1,r_2)=\frac{1}{2\pi i}\frac{\p}{\p j}\log R(j|r_1,r_2)$. 
The reflection amplitude describes the reflection of an incoming plane
wave of spin $j$ coming from spatial infinity, into an outgoing wave
with a phase $R(j|r_1,r_2)$; so $R(j|r_1,r_2)$ comes with
the pysical requirement that its modulus should be one, expressing the
unitarity of the reflection process. This is equivalent to
$N(j|r_1,r_2)$ being real.

The definition of the 
reflection amplitude holds for general quantum mechanical systems
living on noncompact spaces, so that we can define asymptotic states
and study their reflection properties. In general the density of
states suffers from a universal large volume divergence, which can be
regularized by considering
relative reflection amplitudes and densities of states. In our
case this means that we will in fact consider 
$\frac{R(j|r_1,r_2)}{R(j|0,0)}$ (we will sometimes keep this
regularization implicit, as we already did in the above relation
between $R(j|r_1,r_2)$ and $N(j|r_1,r_2)$). 

The consistency condition deriving from factorization constraints was
found in \cite{pst}(formula (4.43))
\bea
\frac{R(j+\half|r_1,r_2)}{R(j-\half|r_1,r_2)}
=\frac{2j}{2j+1}e_-(-j-\half|r_1,r_2),\ \ \forall j\in -\half+i\R,
\label{good} \eea
where we use the function 
\bea e_-(j|r_1,r_2)&=&\frac{\Gamma(1+b^2(2j-1))\Gamma(-b^2(2j+1))}{\sin
  \pi b^22j}\ \times \nonumber \\
&\times &\Pi_{s=\pm}\cos(\pi b^2j+s\frac{i}{2}(r_1+r_2))\sin(\pi
b^2j+s\frac{i}{2}(r_1-r_2)),
\eea
where $b^2$ is related to the level $k$ by $b^2=\frac{1}{k-2}$, and we
omitted a $b$-dependent factor. Note that these quantities were
originally written in terms of parameters $\rho_{1,2}$ instead of
$r_{1,2}$, but solving the case $r_1=r_2$ was enough to determine
$\rho$ as a function of $r$ and $b$. Moreover, note that the
consistency condition (\ref{good}) involves the analytic continuation of
the amplitude $R(j|r_1,r_2)$ outside the physical range. This
continuation will give rise to subtleties when $r_1\neq r_2$. 

Now let me solve the equation (\ref{good}).
First recall the solution when $r_1$=$r_2$, already found in \cite{pst}:
\bea R(P|r,r)=\frac{S_k(\rr+P)}{S_k(\rr-P)},\label{same}\eea
here we omit $r$-independent factors and use the function $S_k$
\bea 
\log S_k(x)=i\int_0^\infty\frac{dt}{t}\left(\frac{\sin
    2tb^2x}{2\sinh b^2t\sinh t}-\frac{x}{t}\right). \label{fct}
\eea
For $R(j|r,-r)$, let me write an ansatz obtained 
by replacing $S_k$ with a new function $S'_k$ in eq. (\ref{same})~:
\bea 
\log S'_k(x)=i\int_0^\infty\frac{dt}{t}\left(\frac{\cosh t\sin
    2tb^2x}{2\sinh b^2t\sinh t}-\frac{x}{t}\right). 
\label{spk}
\eea
This function $R(P|r,-r)=\frac{S'_k(\rr+P)}{S'_k(\rr-P)}$ 
can also be rewritten
\bea 
R(j|r,-r)=\sqrt{\frac{R(j|r+\ip,r+\ip)R(j|r-\ip,r-\ip)}
  {R(j|\ip,\ip)R(j|-\ip,-\ip)}}R(j|0,0). \label{oppo}
\eea 
Once we know a reflection amplitude $R(j|r,-r)$ satisfying
(\ref{good}), it is easy to find a solution with arbitrary $r_{1,2}$:
\bea
R(j|r_1,r_2)=R(j|\frac{r_1+r_2}{2},\frac{r_1+r_2}{2})
R(j|\frac{r_1-r_2}{2},\frac{r_2-r_1}{2})R(j|0,0)^{-1}. \label{any}
\eea

Our expressions for $R(j|r,-r)$ and $R(j|r_1,r_2)$ are perfectly
well-defined and regular on the physical line $j\in -\half+i\R$. However,
in order to prove that our ansatz (\ref{oppo}) satisfies
eq. (\ref{good}), we have to define the analytic continuations
involved in eq. (\ref{good}) as well as the squareroot in
(\ref{oppo}). First notice that the function $S_k(x)$  is analytic in
the strip $|\Im x|<\frac{1+b^2}{2b^2}$ and satisfies 
\bea
\frac{S_k(x-\halfi)}{S_k(x+\halfi)}=2\cosh\pi b^2x,\label{rel} 
\eea
so it can be continued to a meromorphic function on the whole complex
plane. Similarly, $S'_k(x)$ is analytic in the strip $|\Im P|<\half$. 
In order to evaluate eq. (\ref{good}) we need to go to the boundary
of this strip, where we meet singularities.
Using eq. (\ref{rel}), $R(P|r,-r)^2$ can easily be defined as a
meromorphic function in the whole complex plane, and it satisfies (the
square of) eq. (\ref{good}). However, $R(P|r,-r)$ has
branch cuts due to the square root. Indeed, the function $S'_k(x)^2$
has a pole at $x=\halfi$ and a zero at $x=-\halfi$, as we can see from
eq. (\ref{rel}) and the identity
\bea
S'_k(x)^2=\frac{S_k(x+\halfib)S_k(x-\halfib)}{S_k(\halfib)S_k(-\halfib)}
S_k(0)^2 
\eea
Branch cuts of $S'_k(x)$ originate at this pole and this zero. 
Notice that arguments of the reflection amplitude belonging to the
physical 
line $j=-\half +i\R$ correspond to real values for the argument $x$ of
$S'_k$. 
By definition (\ref{spk}), the latter function is regular on the real
line, thus its branch cuts cannot cross the real line. 

The full determination of the branch cuts of $S'_k$ is linked with 
the interpretation of the factorization constraint
(\ref{good}). Indeed, this
constraint should hold
in the physical range $j\in
-\half+i\R$, whereas for such $j$ the quantity $R(j+\half|r_1,r_2)$
needs not even be defined. This comes from the fact that this
constraint was obtained from the factorization of an unphysical field
$\Phi ^\half$. As a result, we need to use an analytical continuation
of $R(j|r_1,r_2)$ to unphysical regions, but we have to allow the
possibility of meeting branch cuts. However, the
r. h. s. of (\ref{good}) is a meromorphic function on the whole
complex plane. So it is natural to define the analytic continuation of
$R(j|r_1,r_2)$ outside the physical line by requiring the
l. h. s. of (\ref{good}) to be meromorphic. Equivalently, we require
the constraint eq. (\ref{good}) to hold for all $j$s, not only $j\in
-\half +i\R$.
This means in particular
that the two
branch cuts of $S'_k$ starting at $x=\pm \halfi$ have to be
correlated, in order for the l. h. s. of (\ref{good}) not to
have any branch cuts.
We thus want
$S'_k(x+\halfi)$
and $S'_k(x-\halfi)$ to hit branch cuts simultaneously, so the branch cuts of
$S'_k(x)$ originating at $x=\pm\halfi$ 
should be located either 
at $x=\pm\halfi-\R_+$, or at
$x=\pm\halfi+\R_+$. Then 
the function $\frac{R(j+\half|r,-r)}{R(j-\half|r,-r)}$ is meromorphic
and satisfies eq. (\ref{good}). 

To check these results, one can use them to 
evaluate the annulus amplitude, and compare it with the similar
quantity computed from 
the boundary states. We now write this computation,
but for brevity we neglect all the regularization problems because they have
already been dealt with in \cite{pst}. This means for instance that we
use the character $\chi_P(q)=\frac{q^{b^2P^2}}{\eta(\tau)^3}$ for the
continuous representation of spin $j=-\half +iP$, ignoring the
infinite factor coming from the ground state degeneracy. More
generally, we will work modulo numerical and $\tau$-dependant factors,
since the normalizations have already been fixed in \cite{pst}.
Also, we will
not write explicitly the regularization of the volume divergence which
we already mentioned. This regularization consists in using
$N(j|r_1,r_2)-N(j|0,0)$, it allows us to ignore $r_1,r_2$-independant
terms. 

We now compute the one-loop partition function of an open string
stretching between the two D-branes of parameters $r_1$ and $r_2$,
\bea
Z_{r_1,r_2}(\tq)
&=&\int dP\ N(P|r_1,r_2) \chi_P(q)
\nn
\\
&=&\int dP\ \chi_P(q)\frac{\p}{\p P}\left(\log
R(\j|\frac{r_1+r_2}{2},\frac{r_1+r_2}{2}) \right.
\nn
\\
&& \left. + \log
R(\j|\frac{r_1-r_2}{2},-\frac{r_1-r_2}{2})   \right) 
\nn
\\
&=& \int dP\ \chi_P(q)\frac{\p}{\p P}\log
\frac{S_k(\frac{r_1+r_2}{2\pi b^2}+P) 
  S'_k(\frac{r_1-r_2}{2\pi b^2}+P)}{S_k(\frac{r_1+r_2}{2\pi b^2}-P)
  S'_k(\frac{r_1-r_2}{2\pi b^2}-P)} 
\nn
\\
&=& \int dP\ \chi_P(q) \int dt\ \frac{\cos 2tb^2P}{\sinh b^2 t\sinh
  t}\left( \cos \frac{r_1+r_2}{\pi}t +\cosh t \cos
  \frac{r_1-r_2}{\pi}t\right) 
\nn
\\
&\stackrel{t=2\pi P'}{=}& \int dP'\ \chi_{P'}(\tq) \left[ \frac{\cosh^2\pi
  P'\cos 2r_1P'\cos 2r_2P'}{\sinh
  2\pi b^2P'\sinh 2\pi P'} \right.
\label{zcyl1}
\\
& & \left. -\frac{\sinh^2\pi P' \sin 2r_1P'\sin 2r_2P'}{\sinh
  2\pi b^2P'\sinh 2\pi P'}\right].
\label{zcyl2}
\eea 
This should be equal to the closed-string cylinder diagram, computed
using the boundary state \cite{pst,pol}
\bea
\llangle \Phi ^j_{n,p}(z)\rrangle_r=
\frac{e ^{-r(2j+1)}+(-1)^ne ^{r(2j+1)}}{\Gamma(1+j+\frac{n}{2})
  \Gamma(1+j-\frac{n}{2})}
\Gamma (1+b^2(2j+1)) \Gamma(2j+1)
\frac{2^{-\frac{3}{4}}b^{-\half}
\nu_b^{j+\half}\delta(p)}{|z-\bz|^{2\Delta_j}}. 
\label{etatbordphijnp}
\eea
Here we used the Fourier transform $\Phi ^j_{n,p}$
of the basic bulk field $\Phi
^j(u)$, defined as $\Phi ^j_{n,p}=\int d^2u e
^{-in\arg(u)}|u|^{-2j-2-ip}\Phi ^j(u)$. Now the cylinder diagram is
\bea
Z^{\rm cylinder }_{r_1,r_2}(\tq)=\int dP'\ \chi_{P'}(\tq)
\sum_{n\in\Z}\int dp\ \llangle \Phi
^{\j'}_{n,p}(\halfi)\rrangle_{r_1}^* \llangle \Phi
  ^{\j'}_{n,p}(\halfi)\rrangle_{r_2}.
\label{zcylads2} 
\eea 
In order to check that $Z^{\rm cylinder}_{r_1,r_2}=Z_{r_1,r_2}$ holds (modulo
the numerical factors and the $r_1,r_2$-independant terms that we
neglect), 
we would need to compute $\llangle \Phi
^{\j'}_{n,p}(\halfi)\rrangle_{r_1}^* \llangle \Phi
  ^{\j'}_{n,p}(\halfi)\rrangle_{r_2}$. It is easy to see that this
  quantity depends on $n$ only through the
  parity of $n$.
More precisely, it corresponds to the term (\ref{zcyl1}) if
  $n$ is even, and to the term (\ref{zcyl2}) if $n$ is odd\footnote{Of
  course, in both cases $\sum_n$ gives an infinite result. This can be
  dealt with by cutting off this sum exactly as in \cite{pst}.}.
 
Let me finally explain why no solution to eq. (\ref{good}) was found
in \cite{pst} when $r_1\neq r_2$.
In \cite{pst}, equation (\ref{good}) was not used as such but rewritten
in the form
\bea |R(j+\half|r_1,r_2)|^2=\frac{2j}{2j+1}e_-(-j-\half|r_1,r_2), 
\label{bad} \eea
which was shown not to admit solutions for $r_1\neq r_2$. This is
simply because the r. h. s. is not a real positive number for all
physical values of $j$ if $r_1\neq r_2$.  
But 
the derivation of eq. (\ref{bad}) from eq. (\ref{good}) implicitly
assumes that $R(\bar{j}|r_1,r_2)=\overline{R(j|r_1,r_2)}$. This does not
hold for the $R(j|r,-r)$ that we defined, given the locations of the
branch cuts, which are not related by $j\rightarrow\bar{j}$.
A way to see that is to notice that on the branch cuts the
argument of the square roots is pure imaginary; and
if the function $\sqrt{z}$ has
 a branch cut $i\R _+$ in the $z$-plane then it does not satisfy
$\overline{\sqrt{z}}=\sqrt{\bar{z}}$.

\vskip 0.2cm
\centerline{\bf Acknowledgements}
\vskip 0.1cm
\noindent
I am very grateful to V. Schomerus and J. Teschner for collaborating on related
issues, discussing the present results, and making helpful comments on
this manuscript. I thank the AEI, Potdam for
hospitality 
when this research was done.

%%%%%%%%%%%%%%%%%%%%%%%%%%%%%%%%%%%%%%%%%%%%%%%%%%%%%%%%%%%%%%%%%%%%%%%%

%Bibliography

\end{document}